\documentstyle[eqsecnum,aps]{revtex}

\begin{document}
\title{Many-Body Diffusion and Path Integrals for Identical Particles}
\author{L. F. Lemmens}
\address{Departement Natuurkunde, Universiteit Antwerpen (RUCA),\\
Groenenborgerlaan 171, B-2020 Antwerpen}
\author{F.~Brosens\cite{Author1} and J.T. Devreese\cite{Author2}}
\address{Departement Natuurkunde, Universiteit Antwerpen (UIA),\\
Universiteitsplein 1, B-2610 Antwerpen}
\date{October 26, 1995. Published: Phys. Rev. {\bf E 53}, {\bf 4467} (1996)}
\maketitle

\begin{abstract}
For distinguishable particles it is well known that Brownian motion and a
Feynman-Kac functional can be used to calculate the path integral (for
imaginary times) for a general class of scalar potentials. In order to treat
identical particles, we exploit the fact that this method separates the
problem of the potential, dealt with by the Feynman-Kac functional, from the
process which gives sample paths of a non-interacting system. For motion in
1 dimension, we emphasize that the permutation symmetry of the identical
particles completely determines the domain of Brownian motion and the
appropriate boundary conditions:\ absorption for fermions, reflection for
bosons. Further analysis of the sample paths for motion in 3 dimensions
allows us to decompose these paths into a superposition of 1-dimensional
sample paths. This reduction expresses the propagator (and consequently the
energy and other thermodynamical quantities) in terms of well-behaved
1-dimensional fermion and boson diffusion processes and the Feynman-Kac
functional.
\end{abstract}

\pacs{05.30.Fk, 03.65.Ca, 02.50.Ga, 02.70.Lq}

\section{Introduction}

In the present paper we introduce a new process that combines
multi-dimensional Brownian motion on domains with appropriate boundary
conditions to solve the many-body problem of fermions or bosons interacting
through a general class of scalar potentials. This new process is a
superposition of orthogonal and independent fermion and boson diffusion
processes combined in a precise and prescribed way. In combination with the
Feynman-Kac functional this approach allows to write the propagator of the
many-body Schr\"{o}dinger equation as an expectation over the functional
along the sample paths generated by the new process. Our method extends the
well known techniques for quantum models with distinguishable particles \cite
{Feynman,FeynHibbs,Feynman2,Durrett,Roepstorff,Schulman} to quantum models
with identical particles, in such a way that the so-called ``sign problem''
\cite{Ceperley,Ceperley2,Klein,Reynolds} for fermions is solved, and that
for bosons and fermions sample paths over configurations \cite
{Feynman3,Wiegel,PollCep,CepePol,LyabVor,Ceperley2} generated by the
permutation symmetry are avoided. This opens the perspective that with the
proper algorithms the new process would improve the standard approach \cite
{JordFos,BernThi,TakaIma,Kalos,Newman,Hall,Alavi} used in path integral
Monte Carlo for fermions as well as for bosons.

Fermion diffusion and boson diffusion are (multidimensional) Brownian
motions on a $n$-dimensional domain $D_n$. If $x_1,x_2,\cdot \cdot \cdot
,x_n $ denote the possible components of the positions of the particles on a
line, the domain $D_n$ is defined by the condition $x_1\geq x_2\geq \cdot
\cdot \cdot \geq x_n$ \cite{Marshall}. Brownian motion with reflection at
the boundary of $D_n$ has been studied in the context of traffic flow models
\cite{HarrRei,Harrison} and will below be identified with the boson
diffusion process, because it leads indeed to the propagator of $n$
non-interacting bosons on a line. Brownian motion with absorption at the
boundary of $D_n$ has been identified by the present authors \cite{LBD,BDL}
as the process which leads to the propagator of $n$ non-interacting fermions
on a line. We have argued that a 3-dimensional (3D) extension of this
process consisting of fermion diffusion in one direction and standard
Brownian motion in the two other directions is sufficient to obtain the
ground state energy of an interacting 3D fermion system. The present
analysis strongly supports our previous arguments. It should be mentioned
that our investigation was originated by Korzeniowski {\em et al. }\cite
{Korzeniowski} and the subsequent discussions on this work \cite
{Foulkes,Caffarel2,Korzeniowski2}.

Because of the fact that for 1D problems the stochastic approach to the
many-body problem for identical particles could be formulated in terms of a
fermion or a boson diffusion process, we concentrated our attention to the
reduction of the sample paths of a 3D process to 3 one-dimensional
processes. For the multi-dimensional Brownian motion of distinguishable
particles this reduction is trivial. Indeed, on the level of the process
each 3D sample path is made up by 1D Brownian motions of the components of
the process vector. For identical particles, such a reduction is not
obvious. But by realizing that the knowledge of the propagator of a
non-interacting many-body system of identical particles suffices to study
how this reduction to independent 1D processes can be performed, the
character of the many-body diffusion could be revealed. The analysis of this
reduction is postponed until section 4 and constitutes the main result of
the present paper. Before studying this reduction, we first summarize in
section 2 the basic underlying concepts, borrowed from the stochastic
approach to the quantum theory of distinguishable particles. In section 3
the fermion and boson diffusion processes are derived and discussed in some
detail.

\section{The Feynman-Kac Functional for distinguishable particles}

In this section we briefly describe how the Feynman-Kac functional and
multidimensional Brownian motion can be used to solve the Schr\"{o}dinger
equation or to obtain the partition function of a many-body problem with
distinguishable particles. The basic equations and the mathematical
notations are presented.

\subsection{The 3D configuration space}

For a {\bf free} particle moving in a $3D$-space it is well known that the
Schr\"{o}dinger equation becomes a diffusion equation if the real time
variable is transformed into an imaginary time variable. This means that the
motion of the particle can be represented \cite{Durrett} by a process $%
\left\{ \vec{X}\left( t\right) ;t\geq 0\right\} $ which can take the
realization $\vec{x}\left( t\right) .$ The probability density relating the
position $\vec{x}\left( t\right) $ to a previous position $\vec{x}^{\prime
}\left( t_0\right) $ is Gaussian and given by
\begin{equation}
\rho \left( \vec{x},t;\vec{x}^{\prime },t_0\right) =\left( \frac m{2\pi
\hbar \left( t-t_0\right) }\right) ^{{3/2}}\exp \left( -\frac{m\left( \vec{x}%
-\vec{x}^{\prime }\right) ^2}{2\hbar \left( t-t_0\right) }\right) .
\label{3dNo}
\end{equation}
The process $\left\{ \vec{X}\left( t\right) ;t\geq 0\right\} $ is related to
standard Brownian motion $\left\{ \vec{B}\left( t\right) ;t\geq 0\right\} $
in $3D$ with infinitesimal variance $\sigma ^2=\hbar /m$%
\begin{equation}
d\vec{X}\left( t\right) =\sqrt{\frac \hbar m}d\vec{B}\left( t\right) .
\label{process}
\end{equation}

If the particle moves in a potential $V\left( \vec{x}\right) ,$ the solution
of the Schr\"{o}dinger equation can be obtained as an average over a
Feynman-Kac functional. This average is the expectation over the process (%
\ref{process}). For an arbitrary function $f\left[ \vec{X}\left( t\right)
\right] $ of the process, the expectation is defined as
\begin{equation}
E_{\vec{x}}\left\{ f\left[ \vec{X}\left( s\right) \right] \right\} =\int d%
\vec{x}^{\prime }\rho \left( \vec{x},t+s;\vec{x}^{\prime },t\right) f\left(
\vec{x}^{\prime }\right) .
\end{equation}
Thus the expectation is the average of $f\left[ \vec{X}\right] $ over all
paths that end in the position $\vec{x}$ at time $t+s,$ starting at all
possible positions $\vec{x}^{\prime }$ at time $t.$ If the function to be
averaged is defined by an integral over time, the process is constructed in
such a way that
\begin{equation}
E_{\vec{x}}\left\{ \int_0^tg\left[ \vec{X}\left( s\right) \right] ds\right\}
=\int_0^tE_{\vec{x}}\left\{ g\left[ \vec{X}\left( s\right) \right] \right\}
ds
\end{equation}
From the properties of the expectation and the process it is easy to show
that for a large class of potentials $V\left( \vec{x}\right) $ the solution
of the ``Schr\"{o}dinger equation'' (in imaginary time)
\begin{equation}
\hbar \frac \partial {\partial t}\Psi \left( \vec{x},t\right) =\frac{\hbar ^2%
}{2m}\nabla ^2\Psi \left( \vec{x},t\right) -V\left( \vec{x}\right) \Psi
\left( \vec{x},t\right)  \label{3dSch}
\end{equation}
is given by
\begin{equation}
\Psi \left( \vec{x},t\right) =E_{\vec{x}}\left\{ f\left[ \vec{X}\left(
t\right) \right] \exp \left[ -\frac 1\hbar \int_0^tV\left( \vec{X}\left(
s\right) \right) ds\right] \right\} ,
\end{equation}
where $f\left[ \vec{X}\left( t\right) \right] $ ensures the initial
condition on $\Psi \left( \vec{x},t\right) .$ If one wants to calculate the
propagator of (\ref{3dSch}), this initial condition is
\begin{equation}
\lim_{t\downarrow 0}\Psi \left( \vec{x},t\right) =\delta \left( \vec{x}-\vec{%
x}^{\prime }\right) ,  \label{delta}
\end{equation}
and hence replacing $f\left[ \vec{X}\left( t\right) \right] $ by a $\delta $%
-function \ref{delta}, one obtains the propagator:
\begin{equation}
K\left( \vec{x},t|\vec{x}^{\prime }\right) =E_{\vec{x}}\left\{ f\left[ \vec{X%
}\left( t\right) \right] \exp \left[ -\frac 1\hbar \int_0^tV\left( \vec{X}%
\left( s\right) \right) ds\right] \right\} ,  \label{3dFKf}
\end{equation}
which solves the Schr\"{o}dinger equation (\ref{3dSch}) and satisfies the
initial condition (\ref{delta}).

The Brownian motion (\ref{process}) has independent increments in the $x$-, $%
y$- and $z$-direction. This means that at the level of this process there is
no difference between the sample paths of one particle moving in 3
dimensions, and 3 particles moving each in 1 dimension. At the level of the
potential the difference between three 1D potentials and one 3D potential is
obvious. The 3D character of the potential and hence of the problem can (at
least partly) be transferred to the process by introducing a local drift
vector $\vec{\mu}\left[ \vec{X}\left( t\right) \right] $ in the process
\begin{equation}
d\vec{X}\left( t\right) =\vec{\mu}\left[ \vec{X}\left( t\right) \right] dt+%
\sqrt{\frac \hbar m}d\vec{B}\left( t\right)  \label{drift}
\end{equation}
and the corresponding modification in the Feynman-Kac functional \cite
{Fisher,Caffarel1}. But it is clear that changing the process according to (%
\ref{drift}) only relabels the state space.

\subsection{The $3n$-dimensional configuration space}

The generalization of the preceding stochastic description of the evolution
in the imaginary time domain of a quantum system with $3$ degrees of freedom
to a system with $3n$ degrees of freedom is straightforward. A point of the
configuration space given by a $3n$-dimensional vector $\bar{x}$ represents
now a state of the process. For convenience in the treatment below, we use
the following labeling (for brevity writing down the row vector $\bar{x}^T$)
\begin{equation}
\bar{x}^T=\left(
\begin{array}[t]{llll}
x_1,y_1,z_1, & x_2,y_2,z_2, & \cdots , & x_n,y_n,z_n
\end{array}
\right) .  \label{vector}
\end{equation}
The process $\{\bar{X}\left( t\right) ;t\ge 0\}$ with realization $\bar{x}%
\left( t\right) $ at time $t$ is a configuration obtained according to a
straightforward extension of (\ref{process}):
\begin{equation}
d\bar{X}\left( t\right) =\sqrt{\frac \hbar m}d\bar{B}\left( t\right) ,
\label{3nprocess}
\end{equation}
where $\{\bar{B}\left( t\right) ;t\ge 0\}$ is the $3n$-dimensional Brownian
motion. The transition probability density for an increment $\left( \bar{x}-%
\bar{x}^{\prime }\right) $ in a time lapse $s$ for this process is given by
\begin{equation}
\rho \left( \bar{x},t+s;\bar{x}^{\prime },t\right) =\left( \frac m{2\pi
\hbar s}\right) ^{3n/2}\exp \left( -\frac{m\left[ \bar{x}-\bar{x}^{\prime
}\right] ^T\left[ \bar{x}-\bar{x}^{\prime }\right] }{2\hbar s}\right) .
\label{3ndNo}
\end{equation}

The matrix product in (\ref{3ndNo}) can equally well be written as a dot
product
\begin{equation}
\left[ \bar{x}-\bar{x}^{\prime }\right] ^T\left[ \bar{x}-\bar{x}^{\prime
}\right] =\left( \bar{x}-\bar{x}^{\prime }\right) \cdot \left( \bar{x}-\bar{x%
}^{\prime }\right) =\left( \bar{x}-\bar{x}^{\prime }\right) ^2,
\end{equation}
but the matrix notation will be more advantageous for our purposes.

For the multi-dimensional process $\left\{ \bar{X}\left( t\right) ;t\geq
0\right\} $ the propagator solution of
\begin{equation}
\hbar \frac \partial {\partial t}\Psi \left( \bar{x},t\right) =\left( \frac{%
\hbar ^2}{2m}\sum_{i=1}^{3n}\frac{\partial ^2}{\partial x_i^2}-V\left( \bar{x%
}\right) \right) \Psi \left( \bar{x},t\right)
\end{equation}
can again be written as a Feynman-Kac functional:
\begin{equation}
K\left( \bar{x},t|\bar{x}^{\prime }\right) =E_{\bar{x}}\left\{ f\left[ \bar{X%
}\left( t\right) \right] \exp \left[ -\frac 1\hbar \int_0^tV\left( \bar{X}%
\left( s\right) \right) ds\right] \right\}  \label{3ndFKf}
\end{equation}
if for $f\left[ \bar{X}\left( t\right) \right] $ a $\delta $-function is
taken to impose the initial condition $\lim\limits_{t\downarrow 0}K\left(
\bar{x},t|\bar{x}^{\prime }\right) =\delta \left( \bar{x}-\bar{x}^{\prime
}\right) $.

It is clear that a straightforward extension of (\ref{drift}) with a drift
vector $\bar{\mu}\left( \bar{X}\left( t\right) \right) $ can facilitate the
actual calculations, but the formulation of the problem with or without
drift is equivalent. Therefore also in the case of $3n$ degrees of freedom,
the process (\ref{3nprocess}) that realizes the sample paths for the
Feynman-Kac functional does not distinguish between $n$ particles moving in
3D or $3n$ particles moving in 1D. This situation will be different in the
case of identical particles which will be considered below.

\section{Permutations and Processes}

A striking example of how permutation symmetry can be used to simplify joint
probability distributions can be found in order statistics \cite{Karlin1}.
The basic idea is that for most samples the order in which the values are
measured is irrelevant, which means that any permutation of the observed
values should have the same probability. This constraint makes the joint
probability distribution symmetrically dependent. The dependence can be
lifted by reducing the sample space to ordered sample points: thus for an
observation $x_1,x_2,\ldots ,x_n$ the sample point $x_{(1)},x_{(2)},\ldots
,x_{(n)}$ is considered where the observed values are ordered in such a way
that $x_{(1)}\geq x_{(2)}\geq \cdots \geq x_{(n)}$.

The idea of lifting the dependence implied by the permutation symmetry
through a reduction of the domain has been put forward before. The basic
ingredient of the Bethe Ansatz \cite{Bethe} is precisely the restriction of
the positions on a line in such a way that the positions remain ordered.
This observation has also been used to solve the nodal plane problem for
fermions in one dimension \cite{Klein}. In particle physics the
spatial-temporal distribution of bosons and fermions in beams has been
obtained using this reduction scheme applied on point processes \cite
{BenaMac,Benard,Macchi,DaleVer}. For a diffusion process we applied an
analogous construction to obtain the propagator of interacting fermions \cite
{LBD}. The process used to realize the sample paths was called the ``fermion
diffusion process'' by the present authors. It is a $n$-dimensional
diffusion with the appropriate (absorption) boundary conditions.

We first illustrate the technique for two particles, and summarize the
properties of the fermion diffusion process. Subsequently it will be shown
how this reduction scheme can also be used to construct the $n$-dimensional
boson diffusion process.

\subsection{Two Particles on a Line}

Let $x_1$ and $x_2$ be the coordinates of the first and the second particle
respectively. The configuration space is two-dimensional: $\left(
x_1,x_2\right) \in R^2$. If the particles are identical the configuration $%
\left( x_1,x_2\right) $ and the configuration $\left( x_2,x_1\right) $
should indicate the same state. For fermions with parallel spin, the
anti-symmetry under the interchange of the two particles is taken into
account by the propagator
\begin{eqnarray}
\left\langle x_1,x_2\left| e^{-tH/\hbar }\right| x_1^{\prime },x_2^{\prime
}\right\rangle =
\begin{array}[t]{c}
\rho \left( x_1,t;x_1^{\prime },0\right) \rho \left( x_2,t;x_2^{\prime
},0\right) \\
-\rho \left( x_1,t;x_2^{\prime },0\right) \rho \left( x_1,t;x_2^{\prime
},0\right) .
\end{array}
\label{2eprop}
\end{eqnarray}
We consider this formula for $x_1\geq x_2$ and $x_1^{\prime }\geq
x_2^{\prime }$, thus for the position elements of the state space being $%
\left( x_1,x_2\right) \in D_2$ and $\left( x_1^{\prime },x_2^{\prime
}\right) \in D_2$. The boundary of $D_2$ is defined by $x_1=x_2$ and denoted
by $\partial D_2$.

It should be noted that the usual factor $1/2!$ in front of the propagator
defined over the configuration space does not occur if the motion is
restricted to the domain $D_2.$ Moreover, the propagator on $D_2$ has all
the properties of a transition probability density of a diffusion process
with absorbing boundary conditions: (i) it is positive on $D_2,$ (ii) it
conserves the probability flux if the boundary state is explicitly introduced%
\cite{VanKampen,GihmSko,absorption}, and (iii) it has the semigroup
property. These requirements can be checked by direct calculation. The
positivity of the propagator can also be understood on the basis of a simple
geometrical argument: the distance between $\left( x_1,x_2\right) $ and $%
\left( x_1^{\prime },x_2^{\prime }\right) $ both in $D_2$ is always smaller
than the distance between $\left( x_1,x_2\right) $ and $\left( x_2^{\prime
},x_1^{\prime }\right) $ except at the boundary $\partial D_2$ where both
distances are equal. For a Brownian motion this relation between the
distances implies that the probability density to go from a point $\left(
x_1,x_2\right) $ in $D_2$ to another point $\left( x_1^{\prime },x_2^{\prime
}\right) $ in $D_2$ in a fixed time lapse $t$ is always larger than the
probability density to go from the same point $\left( x_1,x_2\right) $ to
the reflected point $\left( x_2^{\prime },x_1^{\prime }\right) $ outside $%
D_2 $ in the same time lapse.

Having found the diffusion process for two fermions as a two-dimensional
diffusion process on $D_2$ with absorption on the boundary --the so-called
fermion diffusion process-- it is easy to show that similar considerations
hold for two bosons, starting from the propagator
\begin{eqnarray}
\left\langle x_1,x_2\left| e^{-tH/\hbar }\right| x_1^{\prime },x_2^{\prime
}\right\rangle =
\begin{array}[t]{c}
\rho \left( x_1,t;x_1^{\prime },0\right) \rho \left( x_2,t;x_2^{\prime
},0\right) \\
+\rho \left( x_1,t;x_2^{\prime },0\right) \rho \left( x_1,t;x_2^{\prime
},0\right)
\end{array}
.  \label{2bprop}
\end{eqnarray}
Again the propagator (\ref{2bprop}) is a transition probability density of a
two-dimensional diffusion process on $D_2$, but now with reflecting boundary
conditions on $\partial D_2$. Similarly as for fermions, the required
conditions for such a transition probability density are easily verified. A
nice consequence of the restriction of the state space to the domain $D_2$
is that (for both fermions and bosons) the propagator in $D_2$ satisfies the
initial condition
\begin{equation}
\lim_{t\downarrow 0}\left\langle x_1,x_2\left| e^{-tH/\hbar }\right|
x_1^{\prime },x_2^{\prime }\right\rangle =\delta \left( x_1-x_1^{\prime
}\right) \delta \left( x_2-x_2^{\prime }\right) \text{ if }\left\{
\begin{array}{l}
\left( x_1,x_2\right) \in D_2 \\
\left( x_1^{\prime },x_2^{\prime }\right) \in D_2
\end{array}
\right.
\end{equation}

It is clear that (\ref{2eprop}) and (\ref{2bprop}) can be written
respectively as a determinant and a permanent
\begin{equation}
\left\langle x_1,x_2\left| e^{-tH/\hbar }\right| x_1^{\prime },x_2^{\prime
}\right\rangle =\left|
\begin{array}{ll}
\rho \left( x_1,t;x_1^{\prime },0\right) & \rho \left( x_1,t;x_2^{\prime
},0\right) \\
\rho \left( x_1,t;x_2^{\prime },0\right) & \rho \left( x_2,t;x_2^{\prime
},0\right)
\end{array}
\right| _\xi
\end{equation}
where $\xi =+1$ refers to a permanent (bosons) and $\xi =-1$ means a
determinant (fermions). This observation allows to generalize the process
for two identical particles to a process for $n$ identical particles moving
in 1 dimension, because the form of the transition probability density will
automatically take the boundary conditions into account. The fact that the
transition probability density has to be zero at the boundary implies
absorption at the boundary for the fermion problem \cite
{VanKampen,GihmSko,absorption}. For the boson problem the transition
probability density has to be an extremum at the boundary; this implies that
its normal derivative at the boundary is zero and as a consequence the
boundary condition is a reflection \cite{HarrRei,GihmSko,BhatWay}. (Strictly
speaking, the boundary $\partial D_2$ does not belong to the state space $%
D_2 $ for fermions. But since the propagator is zero on this boundary, there
is no point in making the distinction between the state space of fermions
and bosons as long as the conditions are properly taken into account.)

\subsection{The Fermion Diffusion Process}

In the preceding subsection it was found that the state space $D_2$ for two
indistinguishable particles is found by imposing an ordering on the
configuration space, which introduces an additional boundary. The boundary
condition determines the boson or fermion character of the particles because
reflection at the boundary leads to a symmetric propagator in configuration
space, whereas absorption implies an antisymmetric propagator in
configuration space. If one considers $n$ particles, moving freely on a
line, one obtains $\left( n-1\right) $ extra boundary conditions according
to the rule
\begin{equation}
x_1\geq x_2\geq \cdot \cdot \cdot \geq x_n\Longleftrightarrow \left(
x_1,x_2,\cdot \cdot \cdot ,x_n\right) \in D_n  \label{Dn}
\end{equation}
Defining a diffusion process on $D_n$ with the appropriate boundary
condition, we ensure that the Feynman-Kac functional can be used to
incorporate interactions between the particles\cite{Durrett}. Once this is
realized, all attention can be given to the process. This means that one has
to define the transition probability density to go from an element of $D_n$
to another element of $D_n$ in one time lapse. This transition probability
density of course has to satisfy the conditions for a diffusion process in
order to take advantage of the theory developed for multi-dimensional
diffusion on a domain\cite{BhatWay}. Knowing the transition probability
density, one can construct the sample paths using stationary and independent
increments.

For the fermion diffusion process, this transition probability density is a
Slater determinant, the elements of which are the single-particle
propagators \cite{Karlin2}. The transition probability density to go from $%
\bar{x}^{\prime }\in D_n$ to $\bar{x}\in D_n$ in a time interval $t$ is
\begin{equation}
\rho _F\left( \bar{x},t;\bar{x}^{\prime },0\right) =\det \left| \rho \left(
x_i,t;x_j^{\prime },0\right) \right|
\end{equation}
where $\rho \left( x_i,t;x_j^{\prime },0\right) $ is the 1--dimensional
version of the 3-dimensional propagator (\ref{3dNo}). In \cite{LBD} it was
shown that $\rho _F$ is a transition probability density for a Markov
process on $D_n.$ This new process $\left\{ \tilde{X}\left( t\right) ;t\geq
0\right\} $ can be obtained from the $n$-dimensional process $\left\{ \bar{X}%
\left( t\right) ;t\geq 0\right\} $ given by (\ref{3nprocess}) as follows
\begin{equation}
\tilde{X}\left( t\right) =\left\{
\begin{array}{ll}
\bar{X}\left( t\right) & \text{ for }t\leq \tau _{\partial D_n} \\
\bar{X}\left( \tau _{\partial D_n}\right) & \text{ for }t>\tau _{\partial
D_n}
\end{array}
\right. .
\end{equation}
The Markov time $\tau _{\partial D_n}$ is the first exit time of the domain $%
D_n.$ It should be emphasized that the algorithm described in \cite{LBD} to
realize the sample paths is based on the present formulation of the process.
Another realization of this process can be obtained with the rejection
technique \cite{BDL}.

The semigroup property of the transition probability density $\rho _F$
follows by using the resolution of the unity operator $1=\int_{D_n}\left|
\bar{x}\right\rangle \left\langle \bar{x}\right| d\bar{x},$ whereas the
initial condition
\begin{equation}
\lim_{t\downarrow 0}\rho _F\left( \bar{x},t;\bar{x}^{\prime },0\right)
=\delta \left( \bar{x}-\bar{x}^{\prime }\right)
\end{equation}
follows from the corresponding property for the one-dimensional propagators.

\subsection{The Boson Diffusion Process}

Like in the case of fermions on a line, the indistinguishability of bosons
leads to a state space $D_n$ defined in (\ref{Dn}). This means that $\left(
n-1\right) $ boundary conditions have to be given to find a process for
bosons analogous to fermions. The idea that permutation symmetry can imply
an ordering for bosons identical to that for fermions was recently put
forward \cite{CRV} as a consequence of the arbitrariness of the connection
between the statistics of the particles and the algebraic properties of the
second-quantization operators. Accepting the state space $D_n,$ the analysis
of $D_2$ for bosons suggests that reflection at the boundary $\partial D_n$
of $D_n$ leads to a diffusion process that reflects the Bose-Einstein
statistics.

Let $\bar{x}$ and $\bar{y}$ be two elements of $D_n$, and construct the
following permanent
\begin{equation}
\rho _B\left( \bar{x},t;\bar{y},0\right) =\text{perm}\left| \rho \left(
x_i,t;y_j,0\right) \right| .
\end{equation}

It is clear that $\rho _B$ is positive for all $\left( \bar{x},\bar{y}%
\right) $ pairs, and that it also satisfies the required initial condition
\begin{equation}
\lim_{t\downarrow 0}\rho _B\left( \bar{x},t;\bar{y},0\right) =\delta \left(
\bar{x}-\bar{y}\right) .
\end{equation}
Furthermore, in order that $\rho _B$ can be used as a transition probability
density it has to satisfy the conservation of probability and the semigroup
property
\begin{eqnarray}
\int_{D_n}\rho _B\left( \bar{x},t;\bar{y},0\right) d\bar{y} &=&1; \\
\int_{D_n}\rho _B\left( \bar{x},t;\bar{y},0\right) \rho _B\left( \bar{y},s;%
\bar{z},0\right) d\bar{y} &=&\rho _B\left( \bar{x},t+s;\bar{z},0\right) .
\end{eqnarray}

The conservation of probability can be derived using the property that a
permanent is invariant under an interchange of two rows or columns. Hence
\begin{equation}
\int_{D_n}\rho _B\left( \bar{x},t;\bar{y},0\right) d\bar{y}=\int_{D_n}\frac 1%
{n!}\sum_p\rho _B\left( \bar{x},t;\bar{y}_p,0\right) d\bar{y}=\frac 1{n!}%
\int_{R_n}\text{perm}\left| \rho \left( x_i,t;y_j,0\right) \right| d\bar{y}=1
\end{equation}
where use has been made of the fact that $\rho \left( x_i,t;y_j,0\right) $
conserves probability.

The semigroup property follows with an analogous procedure by extending the
integration domain $D_n$ to $R^n$ using the permutation symmetry and
subsequently using the semigroup property of the single-particle propagators
\begin{equation}
\begin{array}{l}
\int_{D_n\strut}\rho _B\left( \bar{x},t;\bar{y},s\right) \rho _B\left( \bar{y%
},s;\bar{z},0\right) d\bar{y}=\frac 1{n!}\int_{D_n}\sum_p\rho _B\left( \bar{x%
},t;\bar{y}_p,s\right) \rho _B\left( \bar{y}_p,s;\bar{z},0\right) d\bar{y}
\\
=\frac 1{n!}\int_{R^n}\text{perm}\left| \rho _B\left( x_i,t;y_j,s\right)
\right| \times \text{perm}\left| \rho _B\left( y_j,s;z_k,0\right) \right| d%
\bar{y}=\text{perm}\left| \rho _B\left( x_i,t;z_k,0\right) \right| .
\end{array}
\end{equation}
In the last step, the semigroup property of the one-particle propagators
gives rise to $n!$ identical contributions.

In order to see how the integration over two permanents leads again to a
permanent, the following argument might be useful. Denote by $\left| \bar{y}%
\right\rangle $ a fully symmetrized solution of the Schr\"{o}dinger equation
for free bosons, properly normalized\cite{Feynman2}. The resolution of unity
is then given by
\begin{equation}
1=\frac 1{n!}\int_{R_n}\left| \bar{y}\right\rangle \left\langle \bar{y}%
\right| d\bar{y}
\end{equation}
Denoting by $H_i^0$ the Hamiltonian for the $i^{\text{th}}$ free particle,
and by $H^0=\sum_{i=1}^nH_i^0$ the Hamiltonian for $n$ free non-interacting
bosons, a diffusion from $\bar{z}\in D_n$ to $\bar{x}\in D_n$ is given by
\begin{equation}
\rho _B\left( \bar{x},t;\bar{z},0\right) =\left\langle \bar{x}\left|
e^{-H^0t/\hbar }\right| \bar{z}\right\rangle =\frac 1{n!}\int_{R_n}\left%
\langle \bar{x}\left| e^{-H^0\left( t-s\right) /\hbar }\right| \bar{y}%
\right\rangle \left\langle \bar{y}\left| e^{-H^0s/\hbar }\right| \bar{z}%
\right\rangle d\bar{y}
\end{equation}
Reduction of all identical contributions to the preceding integral by
permutation symmetry then leads to
\begin{equation}
\rho _B\left( \bar{x},t+s;\bar{z},0\right) =\int_{D_n}\left\langle \bar{x}%
\left| e^{-H^0t/\hbar }\right| \bar{y}\right\rangle \left\langle \bar{y}%
\left| e^{-H^0s/\hbar }\right| \bar{z}\right\rangle d\bar{y}=\int_{D_n}\rho
_B\left( \bar{x},t;\bar{y},0\right) \rho _B\left( \bar{y},s;\bar{z},0\right)
d\bar{y}
\end{equation}

Therefore $\rho _B\left( \bar{x},t;\bar{y},0\right) $ is a transition
probability density to go from $\bar{y}$ to $\bar{x}$ in a time lapse $t$
for a system of non-interacting identical particles with Bose-Einstein
statistics. The boundary conditions for this process are determined by the
behavior of $\rho _B\left( \bar{x},t;\bar{y},0\right) $ at the boundary $%
\partial D_n.$ Because $\bar{\nabla}\rho _B\left( \bar{x},t;\bar{y},0\right)
$ is zero for $\bar{x}\in \partial D_n$, $\rho _B$ satisfies Neumann
boundary conditions, leading to reflection for the process at the boundary
\cite{Durrett,HarrRei,Harrison,VanKampen,GihmSko,absorption}.

In the mathematical literature the relation between the boson diffusion
process and Brownian motion is referred to as the Skohorod equations \cite
{HarrRei,Harrison}. An account on the relation between Brownian motion and
diffusion on a domain with reflecting boundary conditions can be found in
\cite{HarrRei}.

\section{Many Body Diffusion: a process for Identical Particles moving in 3D}

In this section we will use the known free-particle density matrix of $n$
fermions and $n$ bosons moving in a $3n$ dimensional configuration space to
analyze its reduction to propagators on a state space in such a way that
these propagators are transition probability densities for the processes
discussed in the previous section. The starting point of our analysis is the
projection of the density matrix of distinguishable particles on a density
matrix which has the correct symmetry properties under permutation of the
particle positions:
\begin{equation}
\rho _I\left( \bar{x},t;\bar{x}^{\prime },0\right) =\frac 1{n!}\sum_p\xi
^p\rho \left( \bar{x}_p,t;\bar{x}^{\prime },0\right) .
\end{equation}
The projection operator is a weighted average over all elements $p$ of the
permutation group. The weight is the character $\xi ^p$ of the
representation; i.e. $\xi ^p=1$ for bosons, whereas for fermions $\xi
^{p_{+}}=1$ for even permutations $p_{+}$ and $\xi ^{p_{-}}=-1$ for odd
permutations $p_{-}.$

The free-particle density matrix is {\bf not} a transition probability
density: it does not meet the criterion that the process is in a single
state in the limit $t\downarrow 0,$ i.e. $\lim\limits_{t\downarrow 0}\rho
_I\left( \bar{x},t;\bar{x}^{\prime },0\right) $ does not satisfy the
required initial condition $\delta \left( \bar{x}-\bar{x}^{\prime }\right) .$
For fermions there is the additional complication of the sign problem
because $\rho _I\left( \bar{x},t;\bar{x}^{\prime },0\right) $ is negative in
certain regions of the configuration space. These objections against the
interpretation of $\rho _I\left( \bar{x},t;\bar{x}^{\prime },0\right) $ as a
transition probability density in the configuration space also apply for
motions in one dimension. We removed them in the previous section by
restricting the motion to the state space $D_n$ with the appropriate
boundary conditions.

The questions we have to answer for 3D motion are then: what is the
appropriate state space and which boundary conditions have to be applied?
These questions require a further analysis of $\rho _I\left( \bar{x},t;\bar{x%
}^{\prime },0\right) $.

\subsection{The permutation symmetry}

Consider a vector $\bar{x}$ in the configuration space $R^{3n},$ represented
as in (\ref{vector}) with the $x$, $y$ and $z$ component of the $j^{\text{th}%
}$ particle as the $\left( 3j\right) ^{\text{th}}$, $\left( 3j+1\right) ^{%
\text{th}}$ and $\left( 3j+2\right) ^{\text{th}}$ component of $\bar{x}.$
Its permutations $\bar{x}_p$ can be represented as
\begin{equation}
\bar{x}_p=\left[ p\right] \bar{x}
\end{equation}
where $\left[ p\right] $ is a $3n\times 3n$ dimensional matrix with one $%
3\times 3$ identity matrix on each block row and block column, corresponding
to each particle. For instance, for 2 particles $\left[ p\right] $ can take
one of both forms $\left[
\begin{array}{cc}
I & \bigcirc \\
\bigcirc & I
\end{array}
\right] $ or $\left[
\begin{array}{cc}
\bigcirc & I \\
I & \bigcirc
\end{array}
\right] $ with $I=\left[
\begin{array}{lll}
1 & 0 & 0 \\
0 & 1 & 0 \\
0 & 0 & 1
\end{array}
\right] $ and $\bigcirc =\left[
\begin{array}{lll}
0 & 0 & 0 \\
0 & 0 & 0 \\
0 & 0 & 0
\end{array}
\right] $.

The density matrix for the non-interacting identical particles takes the
form
\begin{equation}
\rho _I\left( \bar{x},t;\bar{x}^{\prime },0\right) =\left( \frac m{2\pi
\hbar t}\right) ^{3n/2}\frac 1{n!}\sum_p\xi ^p\exp \left( -\frac m{2\hbar t}%
\left[ \left[ p\right] \bar{x}-\bar{x}^{\prime }\right] ^T\left[ \left[
p\right] \bar{x}-\bar{x}^{\prime }\right] \right)
\end{equation}
which can readily be rewritten as
\begin{equation}
\rho _I\left( \bar{x},t;\bar{x}^{\prime },0\right) =\left( \frac m{2\pi
\hbar t}\right) ^{3n/2}\exp \left( -\frac m{2\hbar t}\left( \bar{x}\cdot
\bar{x}+\bar{x}^{\prime }\cdot \bar{x}^{\prime }\right) \right) \left( \frac %
1{n!}\sum_p\xi ^p\exp \left( \frac m{\hbar t}\left[ p\right] \bar{x}\cdot
\bar{x}^{\prime }\right) \right)
\end{equation}

\subsection{Projection on even permutations}

We now separate the even permutations $\left\{ p_{+}\right\} $ from the odd
permutations $\left\{ p_{-}\right\} ,$ which can be written as $\left\{
p_{-}\right\} =\left\{ rp_{+}\right\} $, where $r$ is an element of the
permutation group which interchanges two particles. Without loss of
generality we take the first and the second particle. (The same element $r$
has to be used for all elements $\left\{ p_{-}\right\} $). Using the fact
that $\xi ^{p_{+}}=1$ for both fermions and bosons, one obtains
\begin{equation}
\sum_p\xi ^p\exp \left( \frac m{\hbar \tau }\left[ p\right] \bar{x}\cdot
\bar{x}^{\prime }\right) =\sum_{p_{+}}\left(
\begin{array}{l}
\exp \left( \frac m{\hbar \tau }\left[ p_{+}\right] \bar{x}\cdot \bar{x}%
^{\prime }\right) \\
+\xi ^r\exp \left( \frac m{\hbar \tau }\left[ r\right] \left[ p_{+}\right]
\bar{x}\cdot \bar{x}^{\prime }\right)
\end{array}
\right)
\end{equation}
where $\left[ r\right] $ is a $3n\times 3n$ matrix whose operation is to
interchange the coordinates of the first and the second particle. Hence $%
\left[ r\right] $ only differs from the identity matrix in the block column
and the block row corresponding to these particles:
\begin{equation}
\left[ r\right] =\left[
\begin{array}{ccccc}
\bigcirc & I & \bigcirc & \cdots & \bigcirc \\
I & \bigcirc & \bigcirc & \cdots & \bigcirc \\
\bigcirc & \bigcirc & I & \cdots & \bigcirc \\
\vdots & \vdots & \vdots & \ddots & \vdots \\
\bigcirc & \bigcirc & \bigcirc & \cdots & I
\end{array}
\right]
\end{equation}

Note that $\xi ^r=-1$ for fermions and $\xi ^r=1$ for bosons. Elementary
algebra then gives
\begin{equation}
\sum_p\xi ^p\exp \left( \frac m{\hbar \tau }\left[ p\right] \bar{x}\cdot
\bar{x}^{\prime }\right) =
\begin{array}[t]{l}
\sum_{p_{+}}\exp \left( \frac 12\frac m{\hbar \tau }\left[ \tilde{I}%
+r\right] \left[ p_{+}\right] \bar{x}\cdot \bar{x}^{\prime }\right) \\
\times \left(
\begin{array}{l}
\exp \left( \frac 12\frac m{\hbar \tau }\left[ \tilde{I}-r\right] \left[
p_{+}\right] \bar{x}\cdot \bar{x}^{\prime }\right) \\
+\xi ^r\exp \left( -\frac 12\frac m{\hbar \tau }\left[ \tilde{I}-r\right]
\left[ p_{+}\right] \bar{x}\cdot \bar{x}^{\prime }\right)
\end{array}
\right)
\end{array}
\end{equation}
where $\tilde{I}$ denotes the $3n\times 3n$ identity matrix (not to be
confused with the $3\times 3$ identity matrix $I$). Since $\left[ \tilde{I}%
+r\right] $ is invariant if any two particles are interchanged, the
permutation symmetry properties of the density matrix are determined by
those of
\[
\left\{
\begin{array}{ll}
\cosh \frac 12\frac m{\hbar \tau }\left[ \tilde{I}-r\right] \left[
p_{+}\right] \bar{x}\cdot \bar{x}^{\prime } & \text{ for bosons} \\
\sinh \frac 12\frac m{\hbar \tau }\left[ \tilde{I}-r\right] \left[
p_{+}\right] \bar{x}\cdot \bar{x}^{\prime } & \text{ for fermions}
\end{array}
\right.
\]

Because
\begin{equation}
\left[ \tilde{I}-r\right] =\left[
\begin{array}{ccccc}
I & -I & \bigcirc & \cdots & \bigcirc \\
I & I & \bigcirc & \cdots & \bigcirc \\
\bigcirc & \bigcirc & \bigcirc & \cdots & \bigcirc \\
\vdots & \vdots & \vdots & \ddots & \vdots \\
\bigcirc & \bigcirc & \bigcirc & \cdots & \bigcirc
\end{array}
\right]
\end{equation}
one readily obtains
\begin{equation}
\left[ \tilde{I}-r\right] \left[ p_{+}\right] \bar{x}\cdot \bar{x}^{\prime
}=\left( \vec{\xi}_1-\vec{\xi}_2\right) \cdot \left( \vec{x}_1^{\prime }-%
\vec{x}_2^{\prime }\right)
\end{equation}
where $\vec{\xi}_1$ and $\vec{\xi}_2$ are the coordinates of the first and
second particle in $\left[ p_{+}\right] \bar{x}$:
\[
\left[ p_{+}\right] \bar{x}=\left(
\begin{array}{c}
\vec{\xi}_1 \\
\vec{\xi}_2 \\
\vdots
\end{array}
\right) =\left(
\begin{array}{c}
\vec{x}_{p_{+},1} \\
\vec{x}_{p_{+},2} \\
\vdots
\end{array}
\right)
\]

The vector $\vec{x}_j$ indicates the usual three-dimensional position
vector, in contrast to $\bar{x}$ which is a vector of dimension $3n,$ as
described above.

\subsection{The parity of $\rho _I\left( \bar{x},t;\bar{x}^{\prime
},0\right) $ and of its components}

The density matrix of $n$ three-dimensional non-interacting identical
particles is given by
\begin{equation}
\rho _I\left( \bar{x},t;\bar{x}^{\prime },0\right) =
\begin{array}[t]{l}
\left( \frac m{2\pi \hbar t}\right) ^{3n/2}e^{\left( -\frac m{2\hbar t}%
\left( \bar{x}\cdot \bar{x}+\bar{x}^{\prime }\cdot \bar{x}^{\prime }\right)
\right) }\frac 1{n!}%
\mathop{\displaystyle \sum }%
\limits_{p_{+}}e^{\frac m{2\hbar \tau }\left[ \tilde{I}+r\right] \left[
p_{+}\right] \bar{x}\cdot \bar{x}^{\prime }} \\
\times \left\{
\begin{array}{ll}
2\cosh \left( \frac 12\frac m{\hbar t}\left( \vec{\xi}_1-\vec{\xi}_2\right)
\cdot \left( \vec{x}_1^{\prime }-\vec{x}_2^{\prime }\right) \right) & \text{
for bosons} \\
2\sinh \left( \frac 12\frac m{\hbar t}\left( \vec{\xi}_1-\vec{\xi}_2\right)
\cdot \left( \vec{x}_1^{\prime }-\vec{x}_2^{\prime }\right) \right) & \text{
for fermions}
\end{array}
\right.
\end{array}
\label{result}
\end{equation}
which is the key result of the present analysis which allows to answer the
questions above on the state space and its boundary conditions. Indeed, the
decompositions
\begin{equation}
\cosh \vec{a}\cdot \vec{b}=
\begin{array}[t]{l}
\cosh a_xb_x\cosh a_yb_y\cosh a_zb_z \\
+\cosh a_xb_x\sinh a_yb_y\sinh a_zb_z \\
+\sinh a_xb_x\cosh a_yb_y\sinh a_zb_z \\
+\sinh a_xb_x\sinh a_yb_y\cosh a_zb_z
\end{array}
\end{equation}
\begin{equation}
\sinh \vec{a}\cdot \vec{b}=
\begin{array}[t]{l}
\sinh a_xb_x\sinh a_yb_y\sinh a_zb_z \\
+\sinh a_xb_x\cosh a_yb_y\cosh a_zb_z \\
+\cosh a_xb_x\sinh a_yb_y\cosh a_zb_z \\
+\cosh a_xb_x\cosh a_yb_y\sinh a_zb_z
\end{array}
\end{equation}
allow to rewrite the density matrix as a sum of 4 terms $\rho _I\left( \bar{x%
},t;\bar{x}^{\prime },0;\ell =0...3\right) $%
\begin{equation}
\rho _I\left( \bar{x},t;\bar{x}^{\prime },0\right) =\sum_{l=0}^3\rho
_I\left( \bar{x},t;\bar{x}^{\prime },0;\ell \right)  \label{rhosum}
\end{equation}
in which by convention we associate the summation indices $\ell $ as follows
with the combinations of given parity with respect to the reflection plane
orthogonal to the indicated direction
\begin{equation}
\begin{tabular}{c}
Parity of $\rho _I\left( \bar{x},t;\bar{x}^{\prime },0;\ell \right) $ for
bosons \\ \hline
\begin{tabular}{c|cccc}
index & $\ell =0$ & $\ell =1$ & $\ell =2$ & $\ell =3$ \\
reflection plane &  &  &  &  \\ \hline
$\perp x$ & even & even & odd & odd \\
$\perp y$ & even & odd & even & odd \\
$\perp z$ & even & odd & odd & even
\end{tabular}
\end{tabular}
\end{equation}
\begin{equation}
\begin{tabular}{c}
Parity of $\rho _I\left( \bar{x},t;\bar{x}^{\prime },0;\ell \right) $ for
fermions \\ \hline
\begin{tabular}{c|cccc}
index & $\ell =0$ & $\ell =1$ & $\ell =2$ & $\ell =3$ \\
reflection plane &  &  &  &  \\ \hline
$\perp x$ & odd & odd & even & even \\
$\perp y$ & odd & even & odd & even \\
$\perp z$ & odd & even & even & odd
\end{tabular}
\end{tabular}
\end{equation}

\subsection{The orthogonality relations}

The difference in symmetry of these contributions $\rho _I\left( \bar{x},t;%
\bar{x}^{\prime },0;\ell \right) $ has important consequences. Consider a
function $f\left( \bar{x}^{\prime \prime }\right) $ which is invariant under
the permutation of the positions of the particles. It then immediately
follows that
\begin{equation}
\begin{array}[t]{l}
\int d\bar{x}^{\prime \prime }\rho _I\left( \bar{x},t;\bar{x}^{\prime \prime
},0\right) f\left( \bar{x}^{\prime \prime }\right) \rho _I\left( \bar{x}%
^{\prime \prime },t;\bar{x}^{\prime },0\right) \\
=%
\mathop{\displaystyle \sum }%
\limits_{\ell ,\ell ^{\prime }=0}^3\int d\bar{x}^{\prime \prime }\rho
_I\left( \bar{x},t;\bar{x}^{\prime \prime },0;\ell \right) f\left( \bar{x}%
^{\prime \prime }\right) \rho _I\left( \bar{x}^{\prime \prime },t;\bar{x}%
^{\prime },0;\ell ^{\prime }\right) \\
=%
\mathop{\displaystyle \sum }%
\limits_{\ell =0}^3\int d\bar{x}^{\prime \prime }\rho _I\left( \bar{x},t;%
\bar{x}^{\prime \prime },0;\ell \right) f\left( \bar{x}^{\prime \prime
}\right) \rho _I\left( \bar{x}^{\prime \prime },t;\bar{x}^{\prime },0;\ell
\right)
\end{array}
\label{orthogonal}
\end{equation}
where use has been made of the different parity of the components of $\rho
_I\left( \bar{x},t;\bar{x}^{\prime },0\right) $ to reduce the double sum
into a single summation.

The important consequence is that it is sufficient to analyze each component
$\rho _I\left( \bar{x},t;\bar{x}^{\prime },0;\ell \right) $ with a given
parity individually with respect to the interchange of the particles. It is
this property which allows to find a diffusion process for each $\rho
_I\left( \bar{x},t;\bar{x}^{\prime },0;\ell \right) $ separately, as
discussed in the next section.

\subsection{The state space for many-body diffusion}

We now analyze each component $\rho _I\left( \bar{x},t;\bar{x}^{\prime
},0;\ell \right) $ of the density matrix separately. For a given value of $%
\ell $ this function, defined on the configuration space, can be obtained
from a transition probability density defined on the state space $%
D_n^3\equiv D_n\otimes D_n\otimes D_n$ because it is a product of the
transition probabilities of three independent processes, each defined on a $%
D_n$. In other words, the property (\ref{orthogonal}) for the density matrix
of free identical particles allows to reduce the configuration space to a
much smaller state space with independent fermion or boson processes in each
direction.

Let $\left\{ \tilde{X}_{F\ell }\left( t\right) ;t\geq 0\right\} $ be the
process that generates the sample paths for fermions moving in 3D. Then this
process is given according to the following rule
\begin{equation}
\begin{array}{ccccc}
& \ell =0 & \ell =1 & \ell =2 & \ell =3 \\
\tilde{X}_{F\ell }\left( t\right) = & \left\{
\begin{array}{l}
\tilde{X}_F\left( t\right) \\
\tilde{Y}_F\left( t\right) \\
\tilde{Z}_F\left( t\right)
\end{array}
\right. & \left\{
\begin{array}{l}
\tilde{X}_F\left( t\right) \\
\tilde{Y}_B\left( t\right) \\
\tilde{Z}_B\left( t\right)
\end{array}
\right. & \left\{
\begin{array}{l}
\tilde{X}_B\left( t\right) \\
\tilde{Y}_F\left( t\right) \\
\tilde{Z}_B\left( t\right)
\end{array}
\right. & \left\{
\begin{array}{l}
\tilde{X}_B\left( t\right) \\
\tilde{Y}_B\left( t\right) \\
\tilde{Z}_F\left( t\right)
\end{array}
\right.
\end{array}
\end{equation}
where $\tilde{X}_F\left( t\right) ,$ $\tilde{Y}_F\left( t\right) $ and $%
\tilde{Z}_F\left( t\right) $ denote fermion diffusion processes in the $x,$ $%
y$ and $z$ direction. Similarly $\tilde{X}_B\left( t\right) ,$ $\tilde{Y}%
_B\left( t\right) $ and $\tilde{Z}_B\left( t\right) $ are boson diffusion
process in the $x,$ $y$ and $z$ direction. For bosons the decomposition is
as follows
\begin{equation}
\begin{array}{ccccc}
& \ell =0 & \ell =1 & \ell =2 & \ell =3 \\
\tilde{X}_{B\ell }\left( t\right) = & \left\{
\begin{array}{l}
\tilde{X}_B\left( t\right) \\
\tilde{Y}_B\left( t\right) \\
\tilde{Z}_B\left( t\right)
\end{array}
\right. & \left\{
\begin{array}{l}
\tilde{X}_B\left( t\right) \\
\tilde{Y}_F\left( t\right) \\
\tilde{Z}_F\left( t\right)
\end{array}
\right. & \left\{
\begin{array}{l}
\tilde{X}_F\left( t\right) \\
\tilde{Y}_B\left( t\right) \\
\tilde{Z}_F\left( t\right)
\end{array}
\right. & \left\{
\begin{array}{l}
\tilde{X}_F\left( t\right) \\
\tilde{Y}_F\left( t\right) \\
\tilde{Z}_B\left( t\right)
\end{array}
\right.
\end{array}
\end{equation}

For example, the fermion case with $\ell =1$ is invariant under even
permutations $\left\{ p_{+}\right\} $ of the particle coordinates.
Furthermore, under $r$ (interchange of two particles) it is antisymmetric in
the $x$-direction and symmetric in the $y$- and $z$-direction. These
symmetry properties allow to restrict the transitions $\bar{x}^{\prime
}\rightarrow \bar{x}$ to a domain $D_n^3\equiv D_n\otimes D_n\otimes D_n$
simultaneously satisfying the conditions
\begin{equation}
\bar{x}\in D_n^3\Longleftrightarrow \left\{
\begin{array}{l}
x_1\geq x_2\geq \cdots \geq x_n \\
y_1\geq y_2\geq \cdots \geq y_n \\
z_1\geq z_2\geq \cdots \geq z_n
\end{array}
\right.
\end{equation}
with the boundary condition that $\rho _I\left( \bar{x},\bar{x}^{\prime
};\tau ;\ell =1\right) $ is zero if during the transition process the
boundary $\partial D_n$ is hit in the $x$-direction, whereas it is symmetric
with respect to the boundary $\partial D_n$ in the $y$- and the $z$%
-direction. This means that a $n$-dimensional fermion diffusion process is
required in the $x$-direction, and a $n$-dimensional boson diffusion process
in both the $y$- and the $z$-direction.

\subsection{The Feynman-Kac functional}

The many-body process $\left\{ \tilde{X}_{F\ell }\left( t\right) ;t\geq
0\right\} $ for fermions or $\left\{ \tilde{X}_{B\ell }\left( t\right)
;t\geq 0\right\} $ for bosons, defined on the state space $D_n^3,$ can be
used in the same way as $\left\{ \bar{X}\left( t\right) ;t\geq 0\right\} $
to take the interaction between the identical particles into account
\begin{equation}
\begin{array}[t]{c}
K_{_{\left( F,B\right) }}\left( \bar{x},t|\bar{x}^{\prime }\right)
=\sum_{\ell =0}^3K_{_{\left( F,B\right) \ell }}\left( \bar{x},t|\bar{x}%
^{\prime }\right) \\
K_{_{\left( F,B\right) \ell }}\left( \bar{x},t|\bar{x}^{\prime }\right) =E_{%
\bar{x}^{\prime }}\left\{ f\left[ \tilde{X}_{_{\left( F,B\right) }\ell
}\left( \tau \right) \right] \exp \left[ -\frac 1\hbar \int_0^tV\left(
\tilde{X}_{_{\left( F,B\right) }\ell }\left( s\right) \right) ds\right]
\right\}
\end{array}
.  \label{interacting}
\end{equation}

Once the propagator of the interacting system on $D_n^3$ is found with this
prescription, the propagator on the configuration space can be obtained from
the permutations of the particle indices. The function $f\left[ \tilde{X}%
_{_{\left( F,B\right) }\ell }\left( \tau \right) \right] $ expresses again
the initial condition. By construction the propagators found on $D_n^3$ have
the correct symmetry with respect to interchanging of the particles. From
the positivity of the expressions $\rho _I\left( \bar{x},t;\bar{x}^{\prime
},0;\ell \right) $ on $D_n^3$ it follows that $K_{_{\left( F,B\right)
}}\left( \bar{x},t|\bar{x}^{\prime }\right) $ is also positive on the same
domain.

\section{Discussion and Conclusions}

The construction of a state space equipped with a diffusion process to
provide sample paths for the Feynman-Kac functional of an interacting system
of identical particles takes into account their statistics by imposing for
each degree of freedom the appropriate boundary conditions. The spin- states
are left out of the picture by assuming that there are no spin-dependent
interactions involved and therefore the spin as an additional degree of
freedom has not to be considered explicitly. Of course the spin degrees of
freedom are implicitly present because two identical particles are only
considered indistinguishable if they are in the same spin state. The
combined process for a particle to diffuse in the configuration space and to
change its spin state would require a combination of diffusion and a point
process \cite{dangJon}. The latter would change the number of
indistinguishable particles, and hence the dimension of the state space. We
have only considered the case with a fixed number of indistinguishable
particles in the many body problem. Given a fixed number of fermions or
bosons that are not allowed to change their spin degrees of freedom, a
process is constructed that takes into account their statistics. All
configurations which only differ from each other by a permutation of the
coordinates of a particle are represented by the same state. For motion in $%
1D$ this leads to a the fermion diffusion process\cite{LBD,BDL} and a boson
diffusion process, on a state space $D_n$ with respectively absorption or
reflection on the boundary $\partial D_n$.

The motion in $3D$ combined with the permutation symmetry for $n$ identical
objects induces reducible representations in the configuration space.
Fortunately the quantum behavior of a non-interacting many body system is
known and could be used to decompose the propagator of such a system into
independent boson and fermion processes for each orthogonal direction of the
motion. This reduction is the main result of our analysis because it allows
to formulate each process in $3D$ as 4 combinations of $3$ independent $1D$
processes with boundary conditions in the appropriate state space. It is the
combination of the boundary conditions on $D_n\otimes D_n\otimes D_n$ that
determines wether the process in $3D$ is suited for fermions or bosons. For
example a $3D$ process for fermions can contain one $1D$ fermion diffusion
process and two $1D$ boson diffusion processes. If one wants to calculate
the ground state energy of interacting fermions the two $1D$ boson processes
may be replaced by a Brownian motion in $2D$ because the ground state energy
of distinguishable particles and the boson ground state energy attain the
same value in this limit. This observation explains why we found that the
process used formerly \cite{Korzeniowski,LBD} to study ground state
properties of interacting fermions leads to the correct ground state energy
\cite{BDL} (except when a special direction is chosen to define the $1D$
state space for fermion diffusion which leads to an excited state for
symmetry reasons). The part of the $3D$ process for fermions that embodies
the three $1D$ fermion diffusion processes will decay faster because the
odds to be absorbed are higher. Therefore the contribution of this process
to ground state properties should be negligible. In the case of bosons one
may expect for the same reason --absorption implies a faster decay-- that
the low temperature properties can be deduced from the $3D$ process that
reduces to three $1D$ boson processes.

The main achievement of the present approach for bosons is that the sample
paths over the configuration space taking into account the permutations \cite
{Feynman3,Ceperley2} are replaced by sample paths arising from 3 independent
$1D$ processes with well defined boundary conditions. For fermions the
significance of our approach is that it provides the solution of the so
called sign problem \cite{Ceperley}. Indeed the variance due to the fact
that the propagators are antisymmetric under interchange of particle
coordinates does not arise in our approach. On the level of functional
integration we treat bosons and fermions with the same method. This is not
the case in formerly proposed approaches based on the commutation rules for
bosons and anticommutation rules for fermions \cite{Roepstorff,Shankar}. Our
approach relies fundamentally on the application of group theoretical
concepts to processes, thereby extending previously developed methods for
distributions \cite{Karlin1} and point processes\cite{DaleVer,Macchi}.

Summarizing we can say that we introduced the state space together with the
process that provides sample paths for the Feynman-Kac functional of a many
body problem with a fixed number of identical particles in a given spin
state. The propagator over the configuration space can be obtained by the
application of permutations to a linear combination of such processes. The
boson and fermion diffusion process are fundamental processes in our
approach and their relation to the standard Brownian motion in a domain is
settled by the boundary conditions: absorption for the fermion diffusion
process and reflection for the boson diffusion process.

\acknowledgments
The authors thank J.M.J. Van Leeuwen and W. van Saarloos for a discussion
and P. Platzman for discussions and correspondence on the fermion diffusion
process. Part of this work has been performed in the framework of the
NFWO-projects No. 2.0093.91, 2.0110.91 (``ALPHA-project'') and G. 0287.95,
and in the framework of the European Community Program Human Capital and
Mobility through contract no. CHRX-CT93-0337. One of the authors (FB)
acknowledges the National Fund for Scientific Research for financial support.

\begin{center}
{\bf Erratum Februari 10, 1997:}
\end{center}

In section 4.D. we stated that the double sum in Eq. (4.16)
\[
\int d\bar{x}^{\prime \prime }\rho _I\left( \bar{x},t;\bar{x}^{\prime \prime
},0\right) f\left( \bar{x}^{\prime \prime }\right) \rho _I\left( \bar{x}%
^{\prime \prime },t;\bar{x}^{\prime },0\right) =%
\mathop{\displaystyle \sum }%
\limits_{\ell ,\ell ^{\prime }=0}^3\int d\bar{x}^{\prime \prime }\rho
_I\left( \bar{x},t;\bar{x}^{\prime \prime },0;\ell \right) f\left( \bar{x}%
^{\prime \prime }\right) \rho _I\left( \bar{x}^{\prime \prime },t;\bar{x}%
^{\prime },0;\ell ^{\prime }\right)
\]
reduces to a single sum if $f\left( \bar{x}^{\prime \prime }\right) $ is
invariant with respect to the permutations of the particle positions, where $%
\bar{x}^{\prime \prime }$ is a $3n$ dimensional position vector of $n$
particles in a three-dimensional space. This should be corrected as follows:
the double sum in Eq. (4.16) reduces to a single sum if $f\left( \bar{x}%
^{\prime \prime }\right) $ is invariant with respect to the permutations of
the components of the particle positions. This restricts the class of scalar
potentials for which the method presented can be applied without
modifications, i.e. without taking into account the transitions between $%
\rho _I\left( \bar{x},t;\bar{x}^{\prime \prime },0;\ell \right) $ and $\rho
_I\left( \bar{x}^{\prime \prime },t;\bar{x}^{\prime },0;\ell ^{\prime
}\right) $ for $\ell \neq \ell ^{\prime }$. Furthermore the identity matrix $%
I$ should be replaced by $-I$ in the second row of the first column of the
matrix in Eq. (4.8).


\begin{references}
\bibitem{delta}  For mathematical rigor, one should replace $f\left[ \vec{X}%
\left( t\right) \right] $ by an event of which the expectation is a $\delta $%
-function.
\bibitem{Jensen}  This implies that the Feynman-Jensen inequality can be
used to estimate the partition function for interacting identical particles.
\end{references}
\end{document}